\documentclass[aps,twocolumn,showpacs]{revtex4-1}
\usepackage{dcolumn}
\usepackage{bm}
\usepackage{amssymb}
\usepackage{amsfonts}
\usepackage{amsmath}
\usepackage{graphicx}

\begin{document}

\title{Mathematical control theory, the immune system, and cancer}
\author{Augusto Gonzalez}
\affiliation{Instituto de Cibernetica, Matematica y Fisica, Calle E 309, Vedado, La Habana, Cuba}

\begin{abstract}
Simple ideas, endowed from the mathematical theory of control, are used in order to analyze in general grounds the human immune system. The general principles are minimization of the pathogen load and economy of resources. They should constrain the parameters describing the immune system. In the simplest linear model, for example, where the response is proportional to the load, the annihilation rate of pathogens in any tissue should be greater than the pathogen's average rate of growth. When nonlinearities are added, a reference value for the number of pathogens is set, and a stability condition emerges, which relates strength of regular threats, barrier height and annihilation rate. The stability condition allows a qualitative comparison between tissues. On the other hand, in cancer immunity, the linear model leads to an expression for the lifetime risk, which accounts for both the effects of carcinogens (endogenous or external) and the immune response. 
\end{abstract}

\maketitle

{\bf Immunity and control.} Human immunity is, as there are practically all physical 
aspects of life, a control process. 
Our body senses the number of pathogens in a tissue and a response is generated, which
reduces the pathogen load.

The system compresses a variety of sensors, signals and effectors at the cellular level \cite{Janeway}.
In the present paper, I would like to present a different perspective, endowed
from the mathematical theory of control \cite{Teocontrol}. The view allows us to stress some general trends
and qualitatively compare the response in different tissues. 

The general principles are very simple: first, the pathogen load should be minimized, and second, the used resources should be
minimal. These principles should constrain the parameters describing the immune system in any tissue.

{\bf Linear model.} Let us consider, for example, a very small intensity threat to a given tissue in an 
adult individual. The resident cells of the immune system will trigger a response to clear 
the infection. These are, basically, resident cells of the innate system \cite{ResidentMacroph}. 
The simplest available model for the response is a linear one:

\begin{equation}
\frac{dP}{d\tau}=\alpha_t f_t +a P- b_t P,
\label{linear}
\end{equation}

\noindent
in which the response is proportional to the threat. $\tau$ is time, 
$P$ is the number of pathogens (in some units), and $a$ its rate of growth, typically $\sim$
1/hour for bacteria \cite{rate}. A freely evolving group of a few streptococci, for example, 
would lead in around 40 hours to a colony greater than the number of cells in the lungs. 

The coefficient $b_t$, on the other hand, is the tissue annihilation rate of pathogens which, 
for consistency, should be greater than $a$ in order that small threats do not transform into
acute health problems in short terms. This condition requires enough number of resident 
immune cells in the tissue. 

Finally, $\alpha_t f_t$ is the rate of entrance of pathogens into the tissue. 
The constant $\alpha_t < 1$ will model barrier or mucosal immunity, that is, the flow of pathogens, 
$f_t$, is partially trapped and cleared by the barrier or mucosa (or both). 

According to Eq. (\ref{linear}), a finite load of pathogens is always
annihilated, irrespective of the total number. This unrealistic situation is corrected in nonlinear models, 
characteristic of self-regulated systems.

{\bf Nonlinear model.} Ref. \onlinecite{PLoS} uses different models in order to describe the time evolution
of infections in patients. We shall use a modification of their Model 5   
in order to take account of nonlinearities:

\begin{equation}
\frac{dP}{d\tau}=\alpha_t f_t +a P \left(1-\frac{P}{P_s}\right)- \frac{b_t P}{1+c_{\infty}P}.
\label{nonlinear}
\end{equation}

\noindent
The $a$ parameter equals 0.6 hours$^{-1}$. The added nonlinearity limits the increase of $P$ to values below 
$P_s=20$, a conventional parameter indicating sepsis. Authors of paper [\onlinecite{PLoS}] use an average value of $b$
for the body of 1.5 hours$^{-1}$, a value satisfying the requirement $b > a$, mentioned in the previous paragraph. 
The parameter $c_{\infty}=5$ limits also the immune response for high values of $P$. 

The main deficiency of Eq. (\ref{nonlinear}) is the lack of a term modeling recruitment of other immune cells. 
However, even from this simple model, we can get important properties with the help 
of the qualitative theory of differential equations \cite{DiffEqs}.

\begin{figure}[ht]
\begin{center}
\includegraphics[width=0.9\linewidth,angle=0]{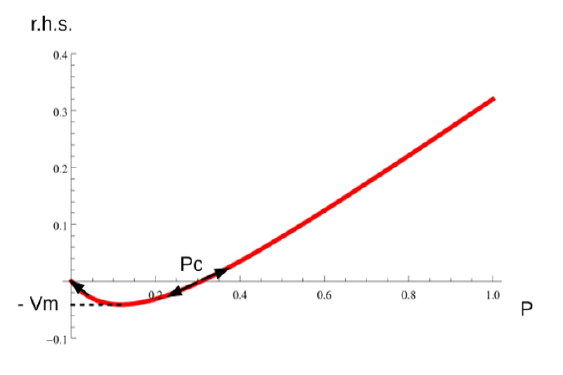}
\caption{The r.h.s. of Eq. (\ref{nonlinear}) for $f_t=0$ and $b_t$=1.5 hours$^{-1}$. A stable fixed point 
at $P=0$, and an unstable one at $P=P_c\approx 0.3$ are signaled. A third, stable one, at $P\approx P_s$,
corresponds to a septic state, which is not seen in the figure. The value of the function at the minimum, $-V_m$,
is indicated.}
\label{fig1}
\end{center}
\end{figure}

{\bf Reference value for the number of pathogens.}
I draw in Fig. \ref{fig1} the r.h.s. of Eq. (\ref{nonlinear}) for $f_t=0$ and $b_t$=1.5 hours$^{-1}$,
which shows two of the fixed points of the equation: $P=0$ (healthy tissue), and $P=P_c$, which is an
unstable fixed point dividing healthy from septic conditions. If, in the time evolution according to
Eq. (\ref{nonlinear}), $P$ reaches values greater than $P_c$, then the final outcome will be a state
with $P$ close to $P_s$. This is the third fixed point of the equation, not seen in the figure.

We can roughly estimate $P_c$ by expanding the r.h.s. of Eq. (\ref{nonlinear}) in series of $P$,
retaining linear and quadratic terms, and equating the result to zero, we get:

\begin{equation}
P_c= \frac{b_t-a}{b_t c_{\infty}-a/P_s}\approx \frac{1}{c_{\infty}}.
\label{Pc}
\end{equation}

\noindent
The last expression comes from neglecting $a$. I will assume that there is a unique $c_{\infty}$
for all of the tissues. This sets {\bf a reference value for $P_c$ in the whole body}. $P_c$ 
could, probably, be associated to the threshold value for initiating recruitment of additional 
immune cells.

{\bf Stability condition in tissues.} 
The response coefficient $b_t$, apart from being greater than $a$, depends on the pathogen load
the tissue is regularly exposed to. If the flow of pathogens, $f$, is practically constant in
certain time intervals, one can get an stability condition 
by requiring the r.h.s. of Eq. (\ref{nonlinear}) to be lower than zero. This leads to:

\begin{equation}
\alpha_t f_t < V_m \approx \frac{b_t}{4 c_{\infty}}\approx \frac{b_t P_c}{4}.
\label{stability}
\end{equation}

\noindent
$V_m$ is the value at the minimum defined in Fig \ref{fig1}. If the inequality is violated, the r.h.s. of 
Eq. (\ref{nonlinear}) is always greater than zero and $P$ increases towards $P_s$. The estimation
for $V_m$ comes from expanding the r.h.s. in series, in the same way as I did for $P_c$.

Coefficients $\alpha_t$ and $b_t$ shall combine in each tissue in order to guarantee Eq. (\ref{stability})
to hold, i.e. guarantee immunity against regular threats. Higher threats would require higher
barriers (smaller $\alpha_t$) and higher annihilation rates ($b_t$). This is typical of epithelial 
tissues. In other cases, for example germinal cells in testis, in order to prevent autoimmunity
the coefficient $b_t$ is reduced, which is compensated by high barriers. In summary, minimization 
of $P$ leads to the condition (\ref{stability})
for the coefficients $\alpha_t$ and $b_t$ in terms of the regular pathogen flow in the tissue, $f_t$.
Economy of resources implies that the inequality should be near optimal. 

\begin{figure}[ht]
\begin{center}
\includegraphics[width=0.9\linewidth,angle=0]{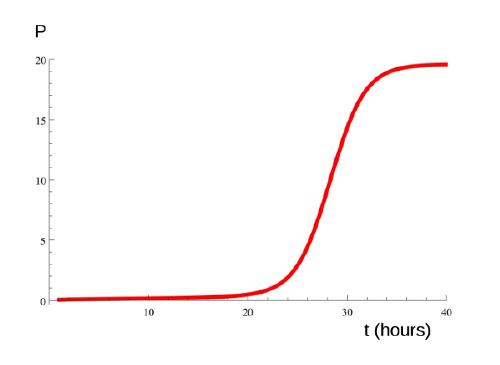}
\caption{Time evolution of $P$ according to Eq. (\ref{nonlinear}) in the unstable regime
$\alpha f=0.05$ hours$^{-1}~ >$ $V_m\approx 0.04$ hours$^{-1}$.}
\label{fig2}
\end{center}
\end{figure}

{\bf A second consequence of the unstable fixed point.}
The unstable fixed point not only sets a unique reference value, $P_c$, but is also the reason for
an interesting property of the small-$P$ response. When the pathogen load overcomes the stability threshold
given by Eq. (\ref{stability}), the fixed point slows down the increase of $P$. The reason is very
simple: $P$ should traverse the region near $P_c$, where the r.h.s. of Eq. (\ref{nonlinear}) is near zero, 
that is, where the annihilation rate of pathogens is close to its rate of growth. 
This is shown in Fig. \ref{fig2}, where $V_m\approx 0.04$ hours$^{-1}$ and $\alpha f=0.05$ hours$^{-1}$. 
The increase of $P$ is delayed for more than 20 hours, in spite of the fact that the 
characteristic time scale of the problem is around one hour. This delay allows recruitment of immune cells
from blood circulation.

{\bf A qualitative comparison.}
The following example is a qualitative comparison between two nearby tissues: the small and large
intestines. An understanding of the reinforced immunity of the small intestine comes from 
this analysis. 

\begin{figure}[ht]
\begin{center}
\includegraphics[width=0.9\linewidth,angle=0]{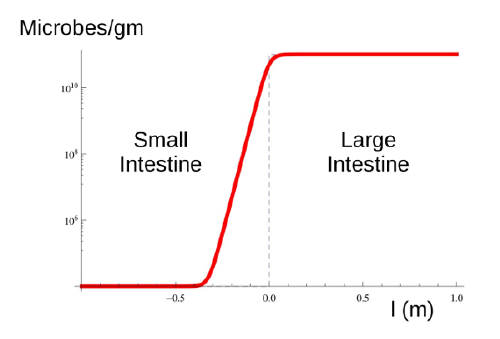}
\caption{Schematic representation of the density of microbes in the gut.}
\label{fig3}
\end{center}
\end{figure}

I show in Fig. \ref{fig3} a schematics of the density of microbes in the
contact region. These microbes are mainly commensal bacteria, but it is reasonable to assume that 
the pathogen loads are proportional to these numbers.

$l$ is a coordinate along the gut. The small bowel is located at $l<0$, and
the large intestine at $l>0$. The mean value of microbes/gm experiences a
jump from $10^4$ to $10^{11}$ as we cross from the ileum to the cecum 
\cite{microbes}. Of course, we expect the dependence to be continuous, as 
schematically represented in Fig. \ref{fig3}.

The parameter values for the large intestine, $\alpha_l$ and $b_l$, are roughly constant. 
In the small intestine, however, the parameters shall exhibit a spatial variation.
$\alpha_s$ shall decrease and $b_s$ increase as $l$ moves towards the distal
end of the ileum. Significant variations of the parameters are expected due
to the augmented flow of pathogens in many orders of magnitude. This is consistent 
with the distribution of Paneth cells \cite{Paneth}, Peyer's patches \cite{Peyer}
and other structures along the small bowel. Above, the immune protection in the 
small intestine was said to be ``reinforced'' in the sense that the coefficients
$\alpha_s$ and $b_s$ shall vary in order to increase protection as $f_t$ increases.  

{\bf Other tissues.} The stability condition, Eq. (\ref{stability}),  allows also
the analysis of tissues in which the flow of pathogens is normal, but $b_t$ 
is decreased at the expenses of lowering $\alpha_t$. These are, for example, the 
brain \cite{Blood-Brain} and testis \cite{Blood-Testis},
where a limit to the cellular immune response is needed for a proper functioning
of the tissue. Recall that the values for $b_t$ can never be lower than $a$, as mentioned. 

In addition, there are also tissues, like the gallbladder, where the microbicide character 
of bile \cite{Bile} could be translated into a lower than average $\alpha_t$ and, possibly, a 
low $b_t$. Notice that we have generalized the meaning of the ``barrier'' 
coefficient, $\alpha_t$, not limited now to anatomical or mucosal barriers.

In conclusion, I assume that the coefficients $\alpha_t$ and $b_t$ take different values for
different tissues. The regular flow the tissue is exposed to, $f_t$, basically determines the 
ratio $b_t/\alpha_t$, according to Eq. (\ref{stability}). The tissue's functioning conditions could 
dictate additional restrictions. For example, in the brain $b_t$ should be relatively low, thus
$\alpha_t$ should be decreased.

{\bf Immunity to cancer.} Although the detailed dynamics of cancer onset and development 
is very complex \cite{Franck}, and partially unknown, one may guess that there should be an 
inverse correlation between the annihilation rate of pathogens in a tissue, $b_t$, and the
risk of cancer. Indeed, the cellular immune response is not only responsible of eliminating 
virus infected cells, for example, but also dysfunctional and precancerous cells in the tissue. 
Thus, a low $b_t$ could be related to a higher than normal cancer risk in that tissue. Conversely, 
one may obtain information for $b_t$ from the frequency distribution of cancer in the body 
tissues.

It is reasonable to assume that, for the initial stages of tumors in a tissue, an equation similar to
Eq. (\ref{linear}) holds: 

\begin{equation}
\frac{dN}{d\tau}=g_c +a_c N- b_c N,
\label{tumors}
\end{equation}

\noindent
where $N$ is the (small) number of precancerous cells, $g_c$ is the rate of creation of such cells
in the tissue, $a_c$ is their division rate, and $b_c$ - the tissue's annihilation rate of dysfunctional
cells. $a_c$ can be estimated from the division rate of stem cells in that tissue, $u_t$, assuming that 
cancer cells originate from stem cells \cite{stem}. Typically, $a_c\sim 1/week$ or even smaller 
\cite{Tomasetti}. On the other hand, $b_c \sim b_t$, as noticed. Thus, $b_c >> a_c$. 

For $g_c$, we may use an equation like $g_c \sim p N_{sc} u_t$, where $N_{sc}$ is the number of 
stem cells, and $p$ - a probability parameter modeling the carcinogenic effect of both internal processes 
(free radicals, for example) or external factors (double strand breaks by ionizing radiation, for example).

\begin{widetext}

\begin{figure}[ht]
\begin{center}
\includegraphics[width=0.9\linewidth,angle=0]{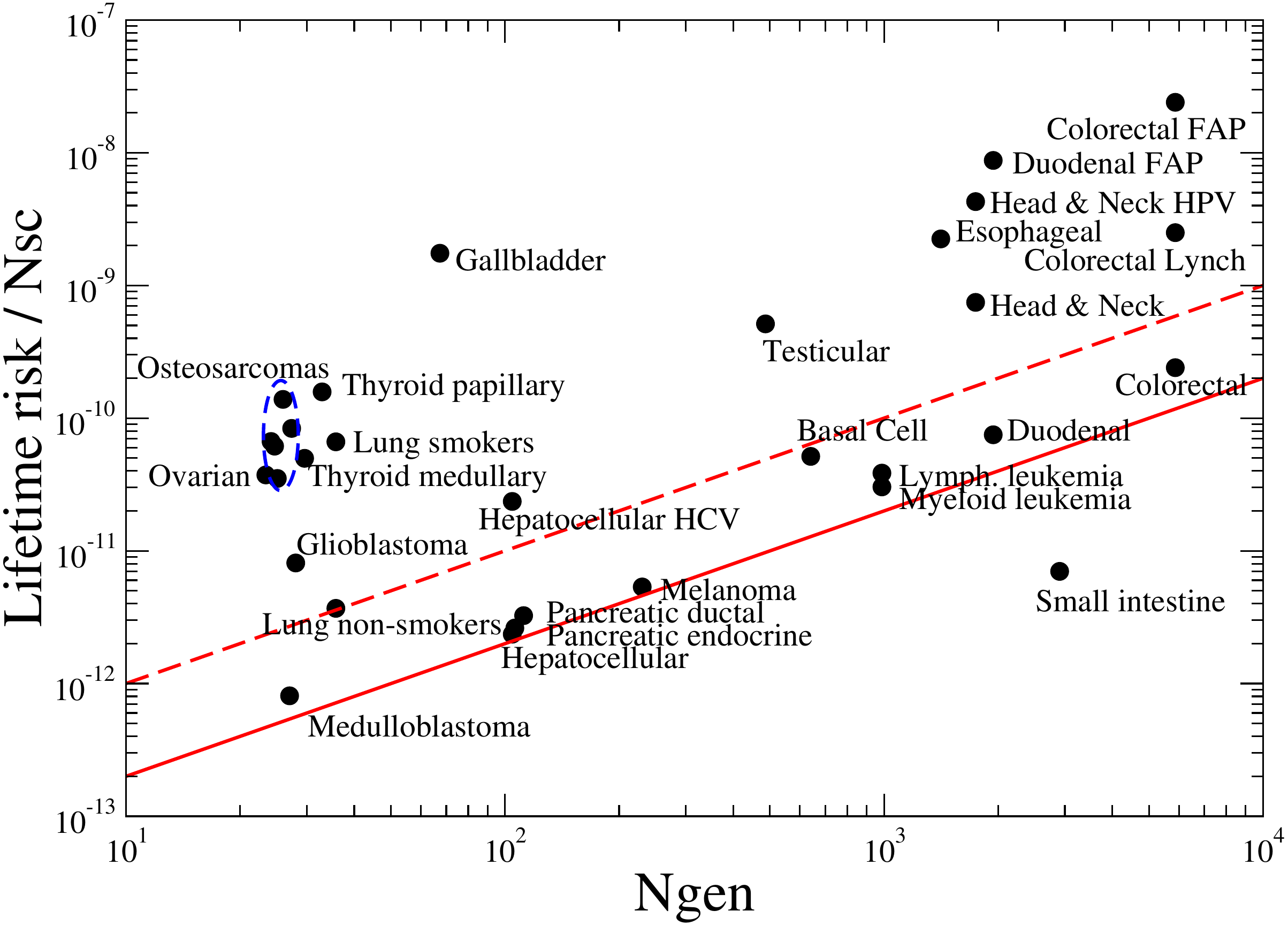}
\caption{Lifetime cancer risk in tissues. The figure is a re-plot of
the data contained in Tomasetti and Vogelstein paper \cite{Tomasetti}.
See explanation in the main text.}
\label{fig4}
\end{center}
\end{figure}

\end{widetext}
\vspace{.5cm}

Equating to zero the r.h.s. of Eq. (\ref{tumors}), we obtain the average number of precancerous
cells in the tissue:

\begin{equation}
N_c\approx g_c/b_c =(p u_t/b_c) N_{sc}.
\label{Nc}
\end{equation}

\noindent
In order to become a true tumor, these cells should pass through a few stages \cite{Franck}, 
and avoid the adaptive immune system. Nevertheless, it is reasonable to assume that the 
lifetime risk for cancer in the tissue is proportional to $N_c$. 

Fig \ref{fig4} is a re-plot of the results by Tomasetti and Vogelstein \cite{Tomasetti} 
(see also \cite{Hannun}) showing the dependence of the lifetime risk for cancer in a tissue on the number of stem
cells, and the rate of mitotic divisions. The $y$-axis
in the figure is the normalized risk, i.e. the risk per stem cell. This normalization allows comparison
between tissues with high differences in the number of stem cells. The $x$-axis, 
on the other hand, counts the number of stem cell generations along lifetime, $N_{gen}$,
a number roughly proportional to the division rate, $N_{gen}\approx u_t~80~{\rm years}+\log_2 N_{sc}$.
The last term accounts for divisions along the clonal expansion phase during tissue formation. 
The figure shows that, for a single-cell lineage, the larger $N_{gen}$ the higher the normalized risk also. 

In Fig. \ref{fig4}, a set of 11 cancers shows a near perfect linear correlation:
$risk/N_{sc}\sim N_{gen}$, where the proportionality coefficient may be roughly written as
$q p/b_t/80$ years, and $q$ measures the success rate of precancerous cells: one in ten thousand cells 
becomes a tumor, for example. 

We shall qualify these tissues as ``normal''. For all
of them, we expect very similar $p$ and $b_t$, although they may exhibit very 
different barriers ($\alpha_t$). Indeed, we expect a very low $\alpha_t$ in the colon
and skin, but $\alpha_t\approx 1$ in blood, for example. The only ``special'' case
in this group is the cerebellum, with a high barrier and a normal (instead of a low) $b_t$. This 
means a possibility higher than the cerebrum to clear any infection \cite{Cerebellum}.

External and genetic factors rise the risk (through $p$) many times, as compared to normal tissues.
For example, smoking multiplies the risk for lung cancer by roughly 20 and, in familial  
adenomatous polyposis patients, the risk for colon cancer is increased by a factor around 100.

There are also tissues like the brain, germ cells, gallbladder, bones and the thyroid
where, in addition to genetic or external factors, the relatively high normalized values for
the risk lead one suspect unusually low values of $b_t$. The brain, germ cells and the gallbladder 
were briefly discussed above. With regard to bones, it is known that immunity relies strongly 
on defensins \cite{Bones}, possibly with a relatively low number of resident cells. On the other hand, 
the thyroid is known to have a close cross-talk with the immune system \cite{Thyroid}. It's
dysregulation is the cause of immune disorders. One may speculate that a low number of
resident cells is needed in order to prevent autoimmunity in the thyroid.

Finally, we have the small intestine with a normalized risk lower than normal, possibly
related to a high averaged $b_t$, a fact consistent with what was discussed above. 

The results for the estimated coefficients are summarized in Table \ref{tab1}. 
Although they are only qualitative results, they allow comparison between tissues, and
are a first step towards the understanding of Fig. \ref{fig4} for the lifetime risks
of cancer in different tissues.

\begin{widetext}

\begin{table}[ht]
\begin{tabular}{|c|c|c|c|}
\hline
 Tissue & Regular pathogen flow ($f_t$) &Barrier height ($1/\alpha_t$) & Annihilation rate ($b_t$)\\
 \hline
 {Small intestine} & Very High & High & High \\
 \hline
 {Colon} & Very High & Very High & Normal \\
 {Lung} & Very High & Very High & Normal \\
 {Skin} & Very High & Very High & Normal \\
 {Duodenum} & High & High & Normal \\
 {Blood} & Normal & Normal & Normal \\
 {Pancreas} & Normal & Normal & Normal \\
 {Liver} & High & High & Normal \\
 {Cerebellum} & Normal & High & Normal \\
 \hline
 {Germ cells} & Normal & High & Low \\
 {Brain} & Normal & High & Low \\
 {Gallbladder} & Normal & High & Low \\
 {Bone} & Normal & High & Low \\
 {Thyroid} & Normal & High & Low \\
 \hline
 {Esophagus} & High & High & Normal \\
 {Head and Neck} & Normal & Normal & Normal \\
 \hline
\end{tabular}
\caption{Immunity in tissues: qualitative comparison.}
\label{tab1}
\end{table}

\end{widetext}

{\bf Concluding remarks.} In the paper, I show that simple ideas, coming from control theory, 
lead to interesting results when applied to the human immune system. 

In the first step, the linear model described by Eq. (\ref{linear}), 
it is shown that stability of a tissue against small pathogen threats requires
$b_t > a$. If this condition is not fulfilled, a small threat would, in a few hours, 
become a serious health problem. 

The linear model is modified, Eq. (\ref{nonlinear}), in order to consider that neither
the number of pathogens nor the response can grow without limits. Healthy ($P=0$) 
and septic ($P=P_s$) states appear as fixed points of the nonlinear (self-regulated)
equation. In the middle of the way, an additional unstable fixed point, $P=P_c$,
signals the transition from healthy to septic regimes. I postulate that $P_c$ is
common to all tissues. At this step, the stability condition is formulated as, 
Eq. (\ref{stability}):

$$\alpha_t f_t < b_t P_c/4.$$

\noindent
this inequality relates the regular flow of pathogens in the tissue, $f_t$, with the
coefficients $\alpha_t$ and $b_t$. We may have, for example, a very high $f_t$ along 
with a very small $\alpha_t$ and a normal $b_t$, as in the colon. Or a normal 
$f_t$, a small $\alpha_t$ and a small $b_t$, as in the brain.

I notice that the distribution frequency of cancer in tissues provides indications
about the strength of $b_t$. This statement follows from an estimation of the
stationary number of precancerous cells in the tissue, which involves both the 
probabilities of carcinogenic factors and the strength of the immune response.
A low $b_t$, as it is presumably the situation in the
gallbladder, for example, would mean a higher than average number of precancerous
cells in the tissue. These cells shall evolve and succeed in avoiding the adaptive 
system in order to give rise to a cancer.

I hope that the model's simplicity and the qualitative results 
following from it will motivate immunologists
to quantify in more precise terms the immune response in tissues and in the whole body. 

{\bf Acknowledgments.}
The author acknowledges support from the National Program 
of Basic Sciences in Cuba, and from the Office of External Activities of the International
Center for Theoretical Physics (ICTP).

\end{document}